\documentclass[runningheads,a4paper]{article}

\usepackage{amssymb}
\usepackage{graphicx}
\usepackage{authblk}

\usepackage{url}

\newcommand{\keywordname}{\textbf{Keywords:}}
\newcommand{\keywords}[1]{\par\addvspace\baselineskip\noindent\keywordname\enspace\ignorespaces#1}

\title{RKappa: Statistical sampling suite for Kappa models}

\author[1,2]{Anatoly Sorokin\thanks{lptolik@icb.psn.ru}}
\author[2]{Oksana Sorokina}
\author[2]{J. Douglas Armstrong}
\affil[1]{Institute of Cell Biophysics RAS, Pushchino, Moscow region, Russia}
\affil[2]{The University of Edinburgh, Edinburgh, UK}

\begin{document}
\maketitle

\begin{abstract}
We present RKappa, a framework for the development and analysis of rule-based models within a mature, statistically empowered R environment. The infrastructure allows model editing, modification, parameter sampling, simulation, statistical analysis and visualisation without leaving the R environment. We demonstrate its effectiveness through its application to Global Sensitivity Analysis, exploring it in ``parallel'' and ``concurrent'' implementations.

The pipeline was designed for high performance computing platforms and aims to facilitate analysis of the behaviour of large-scale systems with limited knowledge of exact mechanisms and respectively sparse availability of parameter values, and is illustrated here with two biological examples.

The package is available on github: https://github.com/lptolik/R4Kappa
\end{abstract}

\section{Introduction}

Dynamic modelling of biological processes is now established as a powerful tool for revealing the systems-level behaviour emerging from the interaction of molecular components. Modelling techniques based on a range mathematical grounds have been introduced over the past century, including kinetic modelling, deterministic and stochastic Petri nets, logical Boolean modelling, etc. The choice of the modelling approach generally depends upon system size, complexity, level of kinetic detail available and expected outcome. However, for a given model building task, there is no guarantee that a sufficient number of parameters are known well enough to approach biological plausibility or to ensure that the resulting simulation will be computationally tractable.

A relatively new modelling approach - rule-based modelling - is one of several developed to deal with combinatorial complexity emerging in multicomponent multistate systems \cite{Chylek}. These have been implemented using several different semantics (Kappa, BioNetGen, StochSim, etc.) and successfully applied to a number of well-described signalling pathways \cite{Danos1,Faeder,Novere}. Rule-based modelling enables representation, simulation and analysis of the behaviour of large-scale systems where knowledge of exact mechanisms and parameters is limited. These features make it very appealing to a wide variety of biological modelling problems \cite{Danos2}.

As an example, a routine task in bioinformatics is the construction of protein-protein interaction networks (PPINs) from a combination of proteomic and interactomic data. PPINs could be naturally extended by applying rule formalism to the protein-protein interactions and inferring the missing quantitative information, thus direct converting static PPI maps into a dynamic model \cite{Sorokina}. The rule-based approach is also an appropriate technique for modelling sophisticated molecular processes such as transcription and translation with highly combinatory mechanisms and relatively limited knowledge for exact kinetic constants \cite{FEBS}.

Rule-based generalisation of many interaction dynamics enables more effective scaling than methods that consider each interaction independently and in detail \cite{Danos2}. An unavoidable drawback for poor defined large-scale systems is overshooting. Local and particularly Global Sensitivity Analysis (GSA) may help resolve these issues by reducing the high-dimensional parameter space into a more tractable number of important parameters that could be measured experimentally \cite{Marino}. 

However, parameter estimation through the use of population-based global optimization techniques and consequent Local and Global Sensitivity Analysis is still a significant drain on computational resources. The combination of larger networks and the opportunity to explore parameter space rapidly demands high-performance computing platforms, such as distributed clusters, parallel supercomputers, etc. 

The process of building and analysing a dynamic model generally consists of the following essential steps: model assembling, model simulation, analysis of the results and model revision. The whole process is highly iterative, therefore, an general-purpose infrastructure that supports all the steps described above would be desirable. 

Indeed, for other widely used modeling techniques, such as ODE solving, a number of effective infrastructures (toolboxes) have been developed and subsequently proven their value such as COPASI, SBTOOLBOX2, SBML-SAT, SBML-PET, PottersWheel, etc \cite{Maiwald,Mendes,Schmidt,Zi1,Zi2}. As an example, SBTOOLBOX2 is based around the SUNDIALS simulating engine and includes a library of Matlab scripts that support model development, model simulation, fitting of models to experimental results, parameter estimation and analysis of results, including the important options for sensitivity and identifiability analysis \cite{Schmidt}. The SBML-SAT toolbox provides the Matlab platform for local and global sensitivity analysis \cite{Zi1}.

For the relatively new rule-based techniques, such infrastructure is sparse in its coverage. For example, a Matlab-based library is available for BioNetGen, enabling parameter scanning, visualization and analysis of simulation results \cite{Sneddon}. 

What is clearly needed is a method that facilitates the development and analysis of rule-based models within a mature statistically empowered framework. Here we present the RKappa package that embodies this need in the widely available statistical package R and demonstrate its effectiveness through its application to Global Sensitivity.

In addition to traditional GSA that we call here ``parallel'' for simplicity, we have introduced a computational experimental setup based upon the distinctive compositionality feature of rule-based models, which was named ``concurrent'' sensitivity. 

We illustrate the concept with two biological models: 1) large interactomic model of postsynaptic density (``parallel'' GSA) and 2) model for transcriptional initiation (``concurrent'' GSA). Presented pipeline for analysis of rule-based models and the statistical evaluation of results was designed for high- performance computing platforms.

\section{Results}

\subsection{Sensitivity modes}

We illustrate our approach to Global Sensitivity Analysis  with Partial Rank Correlation Coefficient (PRCC)  method \cite{Marino}, however it is applicable to eFAST\cite{Marino}, MPSA \cite{Cho} and to any other algorithm that could be splitted into two distinct parts: parameter set evaluation and sensitivity coefficient calculation. 

Presented setup allows running sensitivity analysis in two modes, ``parallel'' and ``concurrent'', depending on model structure and purposes (Figure 1). In standard ``parallel'' sensitivity experiment the parameter definition part is separated from the rule, agent  and observable definition part of the model with following substitution of particular parameter values for each simulation point. In  ``concurrent'' sensitivity experiments, the rule, agent and observable definition part of the model is additionally divided into constant and variable parts, where the constant part does not depend upon parameters varied during sensitivity analysis.

In biology situation when a group of similar molecules could bind another one in concurrent way is not uncommon. For example, transcription factor could bind different parts of DNA with different affinity , or phosphotase could dephosphorylate several substrates with different efficiency. Accordingly, It would be interesting to assess sensitivity of the parameters when more than one element of concurrent group is available and takes part in the interaction. We are able to perform simulation of that kind because of compositionality property of rule based models, when combination of two or more valid models is a valid model itself. Contrary to the traditional approach to the GSA (``parallel''), in ``concurrent'' GSA we create the single model, which consists of models obtained by substitution of parameter values from sampled points in the parameter space in the same way as in ``parallel'' setup, but combined together to form a supermodel. The capability to generate models and simulation jobs for ``concurrent'' GSA is distinctive feature of our pipeline.

\subsection{Pipeline}

We selected R as an appropriate environment for developing a pipeline for several reasons. First there is a wealth of mature and readily available bioinformatics tools developed in R that are directly applicable. These include packages for data integration, analysis and visualisation. Second - R is free and widely available, which allows simple installation and immediate usage. 

We focused on the rule based modelling language Kappa as it has been widely used and extended over the recent years. Several generations of Kappa simulating engines have been developed to the date, where the most recent one - KaSim3 is established as a powerful tool for modeling tasks \cite{KappaLang}. Thus, we aimed to develop a combination of this latest generation of modelling languages, with an effective simulator all embedded in a scalable, statistical framework (R). 

\begin{figure}[ht!]
\centering
\includegraphics[width=\textwidth]{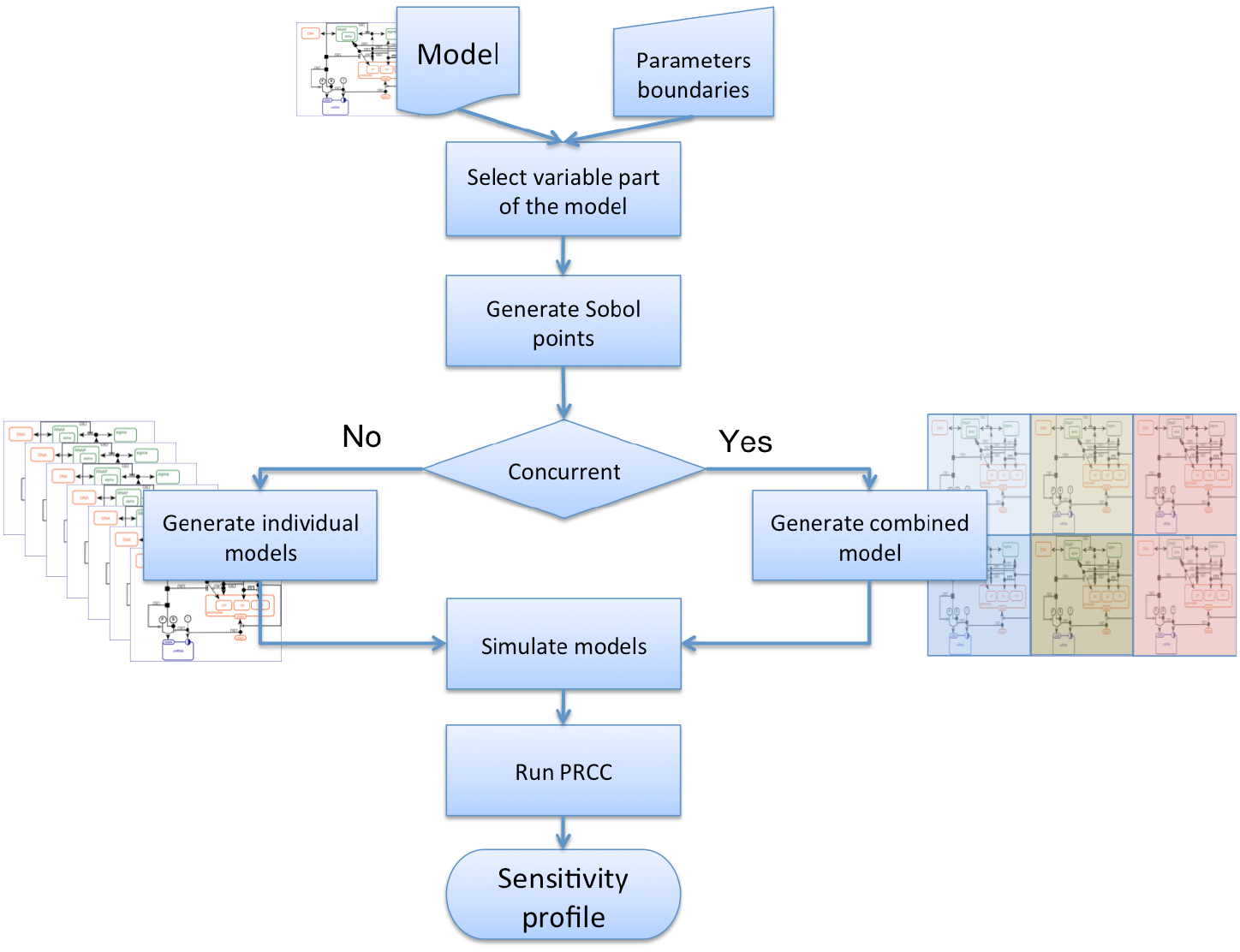}
\caption{RKappa pipeline representation. Once model and ranges of parameter values have been loaded to the pipeline the first step is separation of variable submodel from the constant part, which depends upon parameters of interest.  Next step is the sampling of the parameter space. We are using Sobol low-discrepancy sequence for this, but other methods like latin square are applicable as well. The key step is preparation for the simulation: if ``parallel'' setup is chosen, separate model is generated and simulated for each point in the parameter space independently; for ``concurrent'' setup - variable submodels generated for each point in the parameter space are merged with constant part to form combined model. Generated models are simulated and analyzed as in normal PRCC algorithm.}
\label{fig:pipeline}
\end{figure}

In our pipeline a given Kappa model undergoes the following sequence of processing steps (Figure \ref{fig:pipeline}): 
\begin{enumerate}
\item The model is loaded in R and and getting prepared for further use, for example splitting into separate sections e.g. parameters, initial concentrations, rules, etc.; 
\item The model is modified with respect to the future analysis, for instance, sampling of parameter space can be performed with the Sobol algorithm by specified numbers of points and parameter variation ranges; the corresponding parameter values are then applied to each model instance;
\item Simulation jobs for execution in computer cluster are created and all required data is packaged together.
\item Models are simulated with the appropriate Kappa simulator, such as KaSim \cite{KappaLang};
\item Simulations can be run either locally or with use of parallel computation facilities depending on user demands and task size.
\item Simulation outputs are uploaded to R. Both time series of Kappa ``observables'' and structure of ``snapshots'' \cite{KappaLang} can be analyzed (Figure 2B). ``Snapshot'' structures are converted into iGraph representation for further analysis \cite{igraph}. Graphs are topologically analysed with respect to specified characteristics such as size, composition, ratio of membrane/cytosol elements, etc.,
\item PRCC sensitivity coefficients are calculated by default for each characteristic and the relative parameter impact is visualised with a diagram. Other types of sensitivity metrics could be calculated with `sensitivity' R package if required \cite{Pujol}.
\end{enumerate}
The pipeline may be executed within the R user interface or from the command line. 

The most time-consuming step in any population-based method, either Global Sensitivity Analysis, or parameter fitting, 
is evaluation of the model for particular parameter set. In case when number of parameters is high simulation of tens and even hundreds thousand of parameter sets is required. This necessitates the demand for parallelisation of such algorithms: all parameter sets could be evaluated in parallel, each on its own node. We have implemented this parallelisation 
strategy in our pipeline in a following form: we analyze the model and parameter space locally, generate individual model for each parameter set and prepare the jobs for simulation on high performance computing cluster; the results of the simulation could be analysed locally. 

To be able to generate individual model for each parameter set we split our model into three parts :

\begin{description}
\item[constant part] is the part of the model, which neither depends on nor modified by the application of the new values to the parameter set. In our toy model presented in the package vignette the only statement that belongs to that part is the snapshot definition.
\item[parameter part] is the definition of variables that is substituted by new parameter value assignment %(listing~\ref{code:paramfile}).
\item[template part] is the part that while is not modified by the parameter assignment, influenced by the parameter values directly, like reaction rules, or indirectly, like observables %(listing~\ref{code:varfile}).
\end{description}

In theory for parallel sensitivity we could keep most of the model in constant part, moving to the variable part only reaction rules, which depends upon parameters of interest, but for concurrent sensitivity, when we are going to merge many models into one metamodel, it is important to keep constant part as small as possible. Definition of the same quantity in each sub model could cause syntactic errors when metamodel is formed. 

When models for each parameter set are ready we need to generate scripts to run simulations. Apparently, the pipeline is not tied rigidly to particular simulation engine, instead it requires template for the simulation script to run engine of interest on the cluster. We provide script for KaSim3 engine as default in the code, but user may define its own. The pipeline is able for validation of combination of model and engine locally so syntactic error can be fixed before submission to the calculation cluster. Below there is an example of simulation project creation: 

\medskip
{\it Example of RKappa project creation}
\begin{verbatim}
proj<-prepareProject(project='model',
     numSets=5000,
     exec.path="~/kasim3/KaSim",
     constantfiles=c('model_const.ka'),
     templatefiles=c("model_var.ka"),
     paramfile=c("model_param.ka"),
     type='parallel')
write.kproject(proj)
\end{verbatim}

Here the new project of name ``model'' is created to simulate 5000 parameter sets in ``parallel'' GSA. Last line makes the pipeline to create folder 'model' in the working directory and write everything needed for simulation of all generated models into it. One of the generated KaSim3 simulation scripts is shown below:

\medskip
{\it Example of generated parameter set simulation script}
\begin{verbatim}
#!/bin/bash
numEv=10
time=1000
if [ "$1" != "" ]; then
numEv= $1
echo "number of events to simulate=$numEv" 
fi
if [ "$2" != "" ]; then
time= $2
echo "number of seconsd to simulate=$time" 
fi

i=1
echo $i
mkdir -p "./pset43/try$i"
$KASIM_EXE  -i cABC_const.ka -i param.ka.43 -i cABC_templ.ka.43  \
  -e $time -p 100 -d "./pset43/try$i" -make-sim prom.kasim
while [ $i -lt $numEv ]
do
i=$[$i+1]
mkdir -p "./pset43/try$i"
$KASIM_EXE -e $time -p 100 -d "./pset43/try$i" -load-sim \
  ./pset43/try1/prom.kasim
done
\end{verbatim}
To estimate sensitivity indices of stochastic models like kappa ones, Marino et al \cite{Marino}
proposed to repeat simulation of each parameter set several times and analyse the average of all simulations. In the code above the number of repeated evaluations is defined by parameter ``numEv''.  

As it was said above, generated models could be simulated either locally or remotely by running generated ``job.sh'' script. When simulation is completed the results could be loaded by the command:

\medskip
{\it Example of loading of simulation results}
\begin{verbatim}
abcObs<-read.observables(proj,dir='model')
\end{verbatim}
The most common type of simulation results is observables time course. This type of data is ready for analysis straight after load. For KaSim simulator we built the additional functionality, which makes possible the analysis of the structure of interaction graph or ``snapshot'' obtained at the end of simulation. To perform that type of analysis kappa strings describing complexes created during simulation are converted into igraph subgraphs and combined into final snapshot graph ready for analysis.  
 
We demonstrate our pipeline using two biological model examples as follows. The application of graph metrics for reachability analysis and analysis of parameter sensitivity in "parallel" way is demonstrated by the model of post-synaptic density, while the ``concurrent'' GSA is demonstrated by the transcription initiation model. Both models are available on GitHub as a part of the library documentation.

\subsection{Kappa model of post-synaptic density}

Our first example is a Kappa model of the post - synaptic density (PSD) that was developed to reproduce the core structure of a large (MDa) protein complex underlying the post-synaptic membrane in mammalian neurons. The PSD is believed to mediate the major signal propagation through the synapse and its misfunctioning is thought to underlie many human diseases \cite{Baron}. Proteins of the PSD comprise a wide range of classes including scaffolds, receptors, cytoskeleton proteins and signalling enzymes\cite{Cheng}. They are notably enriched with specific and complimentary domains, such as PDZ and PDZ-binding C-terminal motifs, SH3, GK and some other motifs \cite{Nourry}. This feature was exploited for converting protein-protein interaction into a compacted list of rules based on domain-domain interaction specificity \cite{Sorokina}. The first rule-based models of the PSD described interactions between 50 main structural proteins and reproduced sufficiently the capacity of large protein associations in the post-synaptic compartment, as well as their stability and composition variability \cite{Sorokina}.

An extended model presented here contains 89 proteins and includes signalling (phosphorylation) events in addition to protein association and dissociation processes. As previously, protein-protein interactions are formalised via 543 domain interaction and state modification rules, which require unique 124 parameters to be defined.

For each kinetic constant, values in a biologically sensible range are proposed based on the literature data. We started by sampling the parameter space from a hypercube bound by the dissociation constants (Kd) and rates of dissociation for respective protein interactions. Criteria for selection of required number of simulation points are provided in \cite{Marino} and implemented within the R package. We distributed the massive computation task onto parallel computing facilities, which allowed the simultaneous exploration of 500 models. The key difficulty in simulation of such a big model was the requirement for the steady state reachability. That makes simulation is quite time consuming. In average to simulate 10 repeated evaluation of the parameter set it takes from 10 minutes to 2 hours. Using the Eddie cluster in the Edinburgh University we were able to simulate 500 parameter sets for less then 48 hours.

Upon reaching a steady state, a snapshot of the simulation was collected, parsed into R and the simulation terminated. The obtained protein complexes in the graph representation were processed and analysed with respect to their size, brutto composition, percentage of membrane elements, ratio of membrane/ cytosol proteins in PSD complexes and presence and distribution of surface receptors upstream of key signalling cascades (e.g., NMDA and AMPA receptors). We also calculated the shortest paths between specific members of general signalling cascades resulting in activation of conductivity to understand how closely they are distributed in the generated models. Global sensitivity (``parallel'') was calculated for each of the examined characteristics to evaluate the relative parameter impact.

For each of the 24 specified metrics we also compared ``wild-type'' against four simulated knock-out mutants: PSD95, SAP97, PSD93, SynGap and IRsp53. Figure 2, B shows the example sensitivity diagram for ratio AMPA/NMDA receptors, which is believed to reflect the relative strength of the synapse for the five simulated cases. 

One of the main features of proteins composing the PSD is their functional redundancy, which means that in many cases the complete removal (knock-out) of particular proteins will not completely disrupt key functions such as basal synaptic transmission \cite{Carlisle}. We hypothesised that structural redundancy limits severe changes in size and structure of PSD due to compensation by the remaining proteins. This is exactly what we obtain in model simulations: all the mutated protein complexes except of the most severe case of PSD95 retain the size, ratio of membrane/cytosol proteins and AMPA/NMDA receptors similar to ``wild-type''. 

\begin{figure}[ht!]
\centering
\includegraphics[width=\textwidth]{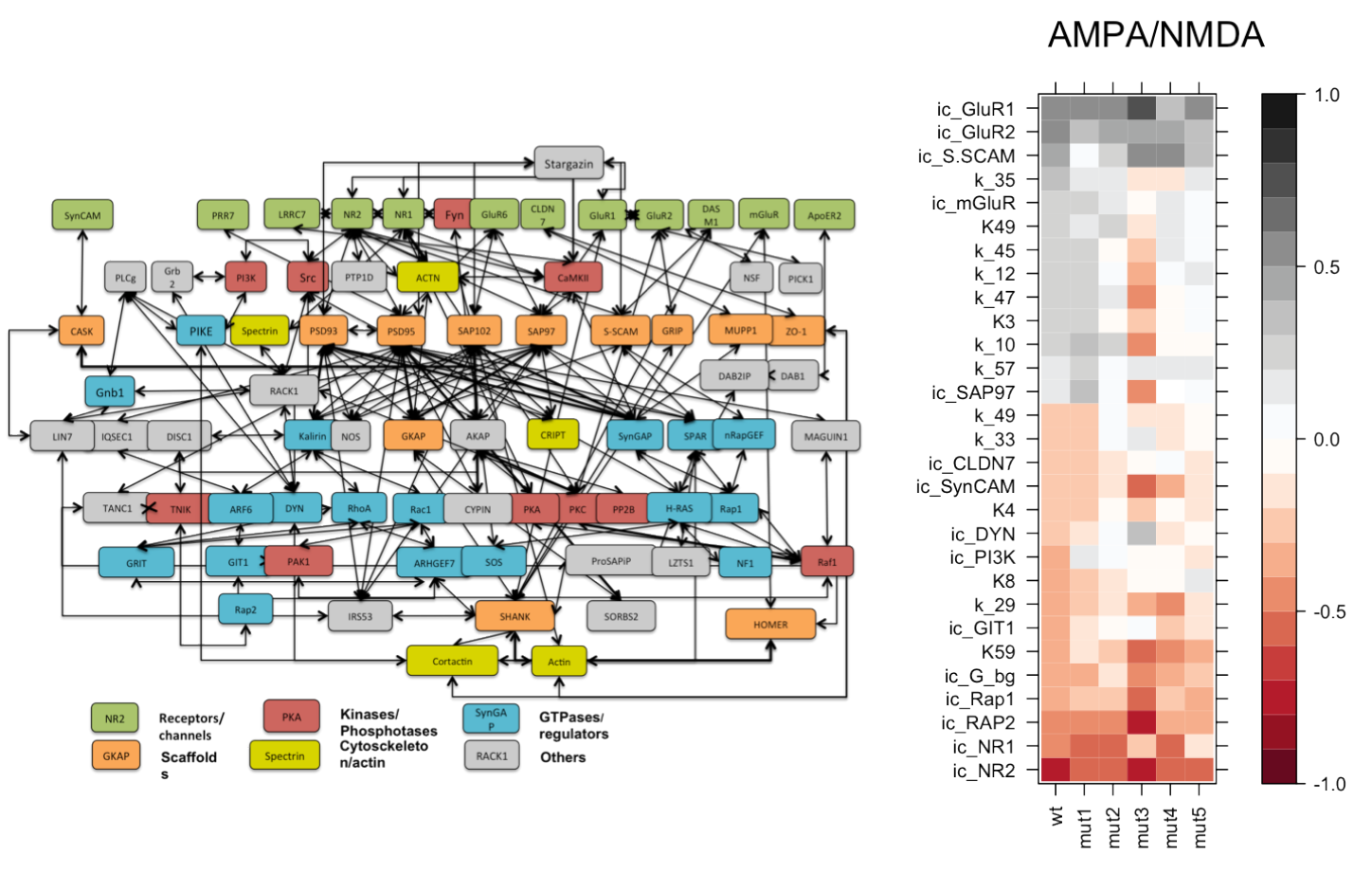}
\caption{Model of post-synaptic density. A. Structure of the model. Different functional categories of proteins are shown in respective colours. B. Results of GSA for AMPA/NMDA ratio in the simulated PSD protein complexes are shown for 6 phenotypes (x-axes). Parameters of binding and unbinding for model components (y-axes) have different impact on the AMPA/NMDA ratio for different phenotypes, which is reflected by colour.}
\label{fig:psd}
\end{figure}

However, in different ``mutant'' phenotypes different kinetic constants appear to respond for the particular characteristic performance (Figure \ref{fig:psd}). Most importantly, length of shortest paths for members of signalling cascades varies between the different mutants and, as that length correlates with signal propagation between nodes \cite{Borgatti}, this in turn is likely to align to some degree with reported electrophysiological abnormalities observed in the in vivo experiments for these mutants.

\subsection{Kappa model of transcription initiation}

\begin{figure}[ht!]
\centering
\includegraphics[width=\textwidth]{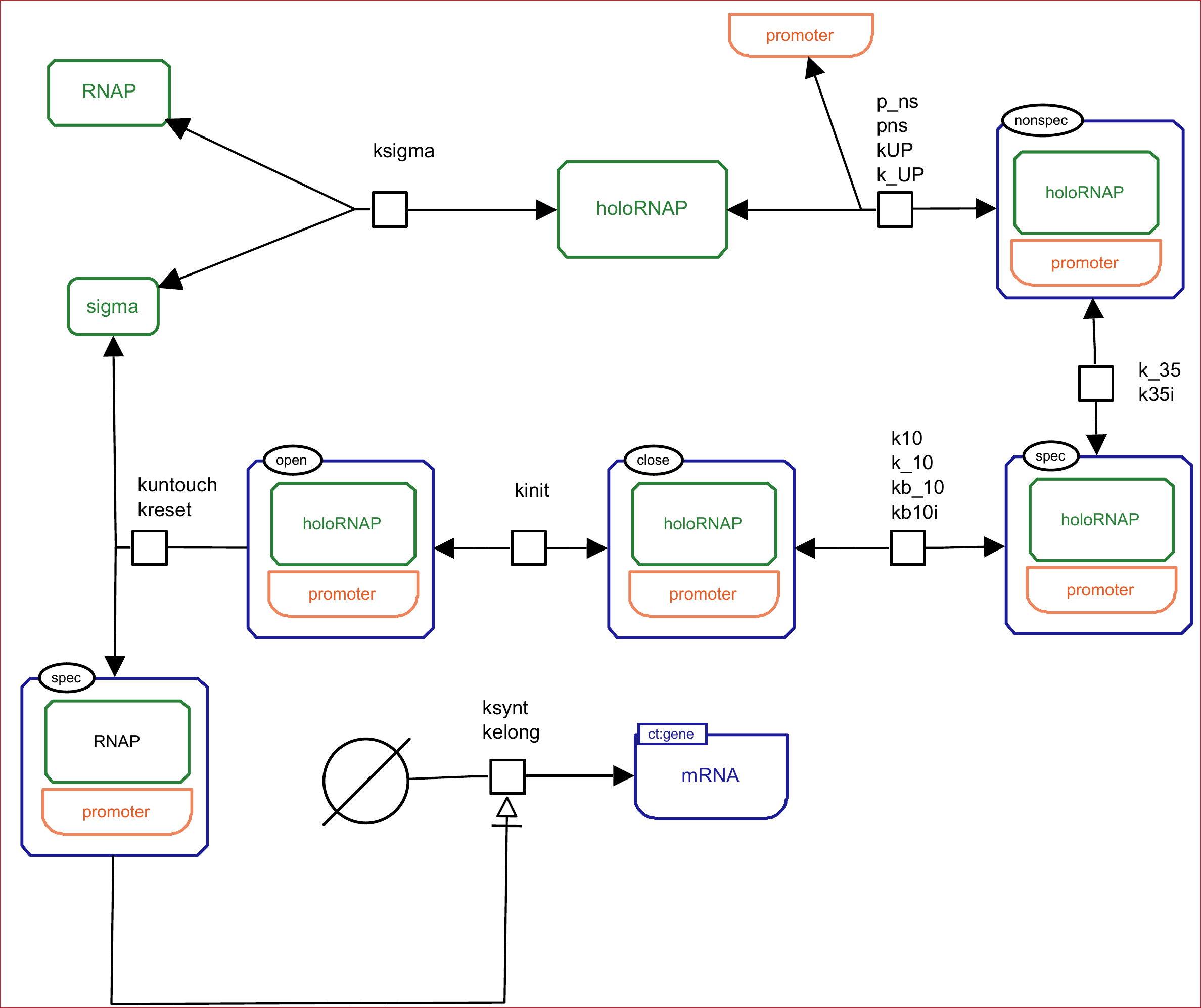}
\caption{Simplified representation of main interactions in the transcription initiation model. Kinetic constants are shown next to reaction nodes.}
\label{fig:sbgn}
\end{figure}

\begin{figure}[ht!]
\centering
\includegraphics[width=\textwidth]{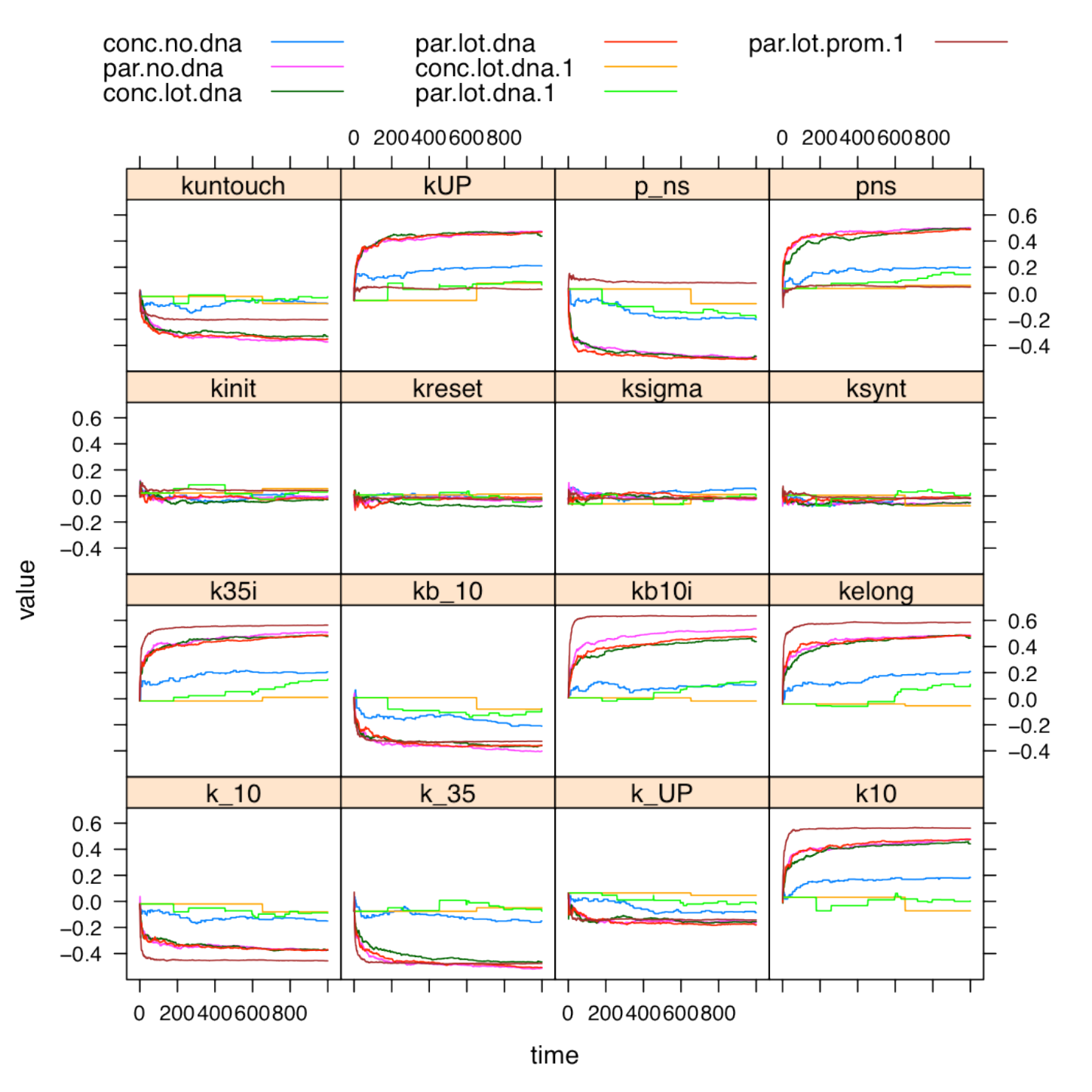}
\caption{Time dependent sensitivity of transcription initiation model at parallel and concurrent conditions. ``Concurrent'' setups: 104 non-specific DNA, 1 promoter, 2000 RNAP (green); 0 no-specific DNA, 1 promoter, 2000 RNAP (blue); 104 non-specific DNA, 1 promoter, 2 RNAP (orange). ``Parallel'' setups: 104 non-specific DNA, 1 promoter, 2000 RNAP (red); 0 no-specific DNA, 1 promoter, 2000 RNAP (magenta); 104 non-specific DNA, 1 promoter, 2 RNAP (yellow); 104 non-specific DNA, 500 promoters, 2 RNAP (pink). It could be seen that PRCC in ``parallel'' and ``concurrent'' setups are quite different, for example ``parallel'' setup without DNA is close to ``concurrent'' setup with a lot of DNA. }
\label{fig:rnap}
\end{figure}

Our second proof-of-concept is a model for bacterial transcription initiation (Figure \ref{fig:rnap}). The Kappa model describes the process of promoter localisation by E. coli RNA polymerase and all five steps of transcription initiation. It was built upon known continuous models and correlates well with in vitro experimental data (\cite{Saecker,Liang}, Supplementary material). It is generally accepted, that the number of available RNA polymerase molecules is smaller (2000-4000 per cell) than number of available promoters (4500-5500), while the number of sigma-subunits that is required by RNA polymerase to bind to the promoter and initiate transcription is even less (about 800 per cell \cite{IshihamaA,IshihamaY}). RNA polymerase, both by itself and in the complex with a sigma-subunit called holoenzyme, is able to bind any site of DNA in a weak nonspecific way, even though it is not able to initiate transcription from it. Therefore, in vivo, promoters have to compete for the active RNA polymerase molecules in the cell. That environment is in stark contrast to the situation generally applied in in vitro experiments to measure various kinetic parameters of RNA polymerase-promoter interaction process: practically all the experimental data used to create current models of transcription initiation are obtained in conditions of the at least equimolar concentration of promoter and RNA polymerase holoenzyme. In some cases concentration of the protein is even three- to ten-fold higher than concentration of the promoter DNA. These environmental differences could conceivably result in significant underestimation of the role of processes such as promoter localisation, initial non-specific binding and the role of non-promoter DNA in transcription initiation. Our kappa model of transcription initiation was developed to allow us to explore the influence of different steps of transcription initiation in competition between various promoters for limited number of holoenzymes in the cell. To prove that the structure of the model is correct we took parameter values from \cite{Sclavi} and show that the simulation results were close to the experimental data.

Here the ``concurrent'' mode is the best choice, as the constant part, which in our case describes RNA-polymerase itself, does not depend upon parameter values varied during sensitivity analysis. The variable part describes interaction between RNA-polymerase and promoters. The supermodel obtained can reflect more accurately the in vivo environment because of the large number of individual promoters characterized by various parameter sets, which allow us to model competition for interaction with limited number of enzymes.

To compare the performance of the ``concurrent'' and ``parallel'' GSA we have run our transcription initiation model with seven different combinations 
of RNAP, promoter and non-specific DNA concentrations. Results of time-dependent sensitivity coefficients for all computational experiments are 
shown on the (Figure 3). We can see that sensitivity profiles of ``parallel'' GSA does not change upon change in amount of non-promoter 
DNA, while in ``concurrent'' GSA most parameters are less sensitive without non-promoter DNA. 

\section{Discussion}

Over the past 10 years, rule-based modelling has established a reputation as a useful tool for molecular simulations. Amongst its benefits, its inherent ability to scale to cope with the high combinatorial complexity of biological systems is arguably the most important.

One example of such system is protein interactomics. Topological analysis of PPI maps is a well-known strategy for learning the basic principles of protein network organization; lots of studies performed up to date to identify the functionally meaningful clusters/motifs in the protein network \cite{Newman,Wang,Pocklington}. 

The rule-based semantics endows a qualitative PPI map with the missing information essential for quantitative modelling; explicitly describing protein binding sites (including possible modifications), concentrations and affinities. Each rule defines what is essential for the particular interaction and omits the irrelevant information, thus, the plethora of concurrent modifying and binding events can be wrapped into a relatively compact executable dynamic model. 

Similar reasoning can be applied to other levels of molecular processes such as transcriptional regulation. Rules describing the essential steps in the interactions between RNA-polymerases and promoters try to take into account as much experimental information as possible whereby an individual plasmid promoter or even all promoters in the cell can potentially be modelled.

The output of any mathematical model is inevitably subjected to uncertainty as the model input is a priori based on several sources of uncertainty, such as absence of information about exact parameter values, erratic measurements and simply poor or partial understanding of the process under investigation. For models built upon the sizable protein-protein interaction networks the number of undefined parameters becomes enormously large and most of them could not be identified experimentally. 

Likewise, it is unlikely possible to define experimentally all the possible rate constants for the binding of individual promotors by RNA-polymerase. So it is essential to understand relative role of transcription initiation steps in the control of gene expression and relate parameters important for in vivo performance to parameters measurable in the in vitro experiments. 

Sensitivity analysis allows us to rank model parameters with respect to their impact within the huge parameter space. GSA in particular allows the global search of the parameter space, changing all the parameter values simultaneously to find the most sensitive subset of parameters [7]. That decrease in task dimensionality makes search for optimal solution much easier.

This approach is based on the assumption that the individual parameters that have biggest impact on the model are also the most important ones to optimise.  Models will obviously struggle to find optimal solutions if they are distributed over a wide range of parameters each of which has minimal individual contribution. However one of the key purposes of this type of modelling is to identify key nodes in molecular networks that can be measured or manipulated in biological systems where the inherent noise would overwhelm such cases.

Kappa evidently lacks a single pipeline for editing the model, configuring initial conditions, iterative modification, simulation, with analysis the results and their visualization. Existing tools do allow running simulations locally, e.g. with KaSim and provide some primary knowledge for model behaviour and structure. These approaches work perfectly well for compact models with well- defined parameters. If one needs to run the multiple versions of large-scale models or compare thousands runs for the given model to explore parameters then existing solutions struggle to cope. Such studies require much more extensive computation and parallel computing becomes a more viable option.

The pipeline we describe here facilitates automatic generation of updated versions of the rule-based models with modified kinetic rates and initial concentrations; it prepares the models for parallel/clustering facilities, simulates them with Kappa simulators (JSim or KaSim), runs GSA and then provides process simulation output with respect to user requirements and finally provides a convenient visualization of the results in a form of graphs. All this can be achieved without leaving R environment, or alternatively from command line.

The current pipeline does not attempt to duplicate the infrastructure designed for building the Kappa models, debugging them and performing initial analysis of their structure as such capabilities are well covered by solutions such as KaSim, simplx/complx, RuleStudio to name a few. Rather we concentrate on managing the simulations and processing of simulation results. The successful implementation of a comprehensive pipeline would naturally entail the design of the infrastructure for the automatic generation the Kappa models from PPi map, Boolean genetic regulatory networks or causality networks inferred from various -omics experiments. However, at the first step using of valid, manually curated models is essential for understanding of the applicability and capacity of the approach.

For both models used as examples the huge parameter space with mostly unknown exact kinetic values makes the task of getting plausible biological insights quite difficult. Global Sensitivity Analysis implemented in two ways allows significant reduction of parameters to the tractable number of most important ones, which in combination with experiments allows model to make realistic predictions.

\subsubsection*{Acknowledgments.} AS was partially supported by RFBR, research project No.  14-44-03679, and European Research Council (ERC) under grants 320823 “RULE”. The research leading to these results has received funding from the European Union Seventh Framework Programme (FP7/2007-2013) under grant agreement nos. 241498 (EUROSPIN project), 242167 (SynSys-project) and 604102 (Human Brain Project). This work has made use of the resources provided by the Edinburgh Compute and Data Facility (ECDF) (http://www.ecdf.ed.ac.uk/). The ECDF is partially supported by the eDIKT initiative (http://www.edikt.org.uk).

\section{The References Section}\label{references}

\end{document}